\documentclass{article}
\usepackage[english]{babel}
\usepackage[utf8]{inputenc}
\usepackage{graphicx}
\usepackage{hyperref}
\usepackage{booktabs}
\usepackage{multirow}
\usepackage{siunitx}
\usepackage[letterpaper,margin=1in]{geometry}
\usepackage[font=small,skip=-1pt]{caption}
\usepackage{amssymb}
\usepackage{amsmath}

\usepackage{hyperref}
\hypersetup{
    colorlinks=true,
    linkcolor=cyan,
    filecolor=cyan,      
    urlcolor=black,
    citecolor=cyan,
}
\usepackage{amsmath}
\usepackage{listings}
\usepackage{float}
\usepackage{multicol}
\usepackage{lipsum}
\usepackage{mwe}

\usepackage{tikz}
\usepackage{tikz-3dplot}
\usepackage{subcaption}
\usepackage{textcomp}

\usepackage[T1]{fontenc}
\usepackage{authblk}

\begin{document}

\title{ \normalsize 
		\LARGE \textbf{Automated Scoring System of HER2 in Pathological Images under the Microscope}
	    }

\author[1,2]{Zichen Zhang}
\author[1]{Lang Wang}
\author[1]{Shuhao Wang}
\affil[1]{Thorough Images, Beijing, China}
\affil[2]{Macalester College, Saint Paul, United States}

\date{}
\maketitle

\begin{abstract} 
{
\hypersetup{linkcolor=black}
Breast cancer is the most common cancer among women worldwide. The human epidermal growth factor receptor 2 (HER2) with immunohistochemical (IHC) is widely used for pathological evaluation to provide the appropriate therapy for patients with breast cancer. However, the deficiency of pathologists and subjective and susceptible to inter-observer variation of visual diagnosis are the main challenges. Recently, with the rapid development of artificial intelligence (AI) in disease diagnosis, several automated HER2 scoring methods using traditional computer vision or machine learning methods indicate the improvement of the HER2 diagnostic accuracy, but the unreasonable interpretation in pathology, as well as the expensive and ethical issues for annotation, make these methods still have a long way to deploy in hospitals to ease pathologists' burden in real. In this paper, we propose a HER2 automated scoring system that strictly follows the HER2 scoring guidelines simulating the real workflow of HER2 scores diagnosis by pathologists. Unlike the previous work, our method considers the positive control of HER2 to make sure the assay performance for each slide, eliminating work for repeated comparison between the current field of view (FOV) and positive control FOV, especially for the borderline cases. Besides, for each selected FOV under the microscope, our system provides real-time HER2 scores analysis and visualizations of the membrane staining intensity and completeness corresponding with the cell classifications. Our rigorous workflow along with the flexible interactive adjustion in demand substantially assists pathologists to finish the HER2 diagnosis faster and improves the robustness and accuracy. The proposed system will be embedded in our $\textit{Thorough Eye}\textsuperscript \textregistered$ platform for deployment in hospitals.
}
\end{abstract}

\section{Introduction}

\begin{figure}[!htb]
\minipage{0.33\textwidth}
  \includegraphics[width=\linewidth]{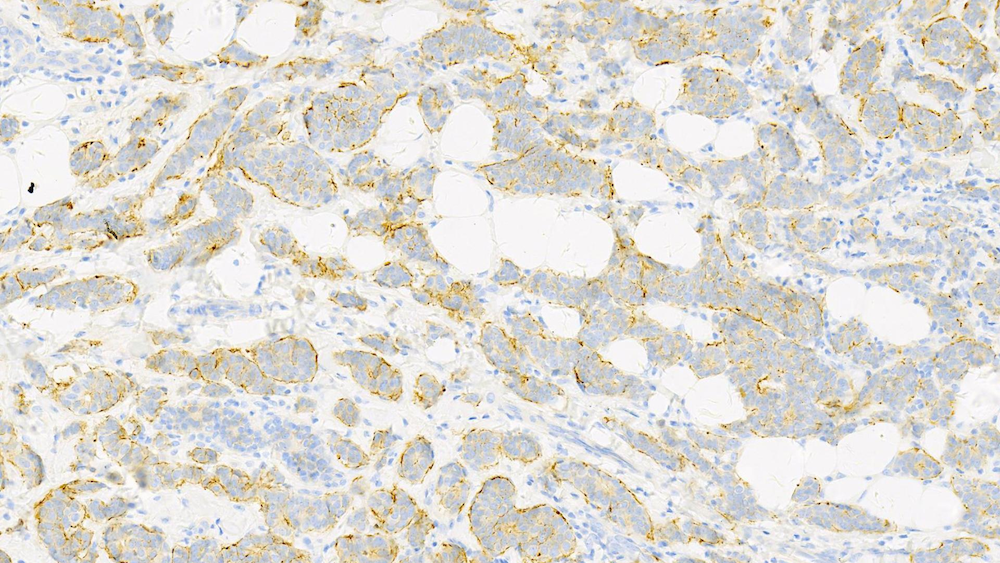}
\endminipage\hfill
\minipage{0.33\textwidth}
  \includegraphics[width=\linewidth]{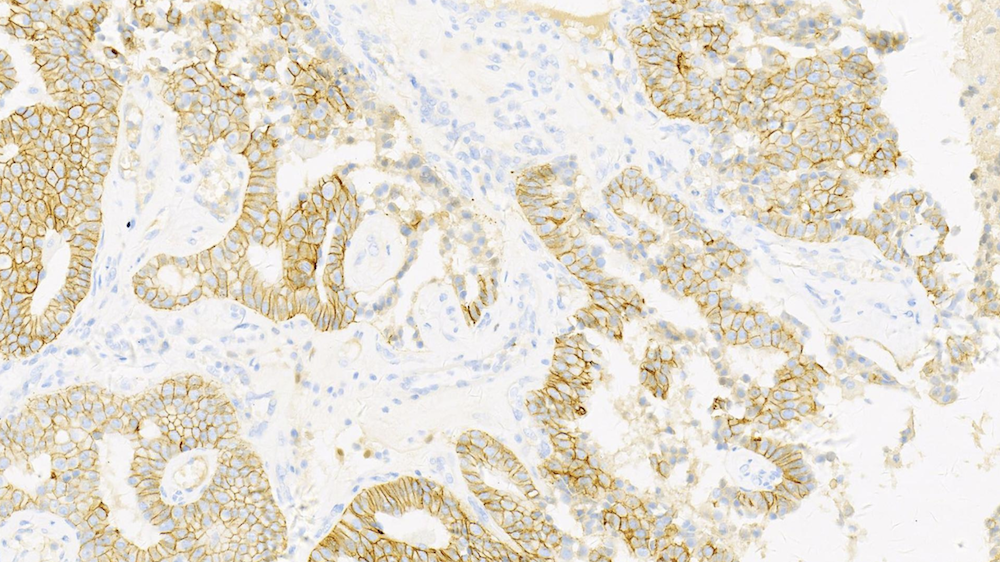}
\endminipage\hfill
\minipage{0.33\textwidth}%
  \includegraphics[width=\linewidth]{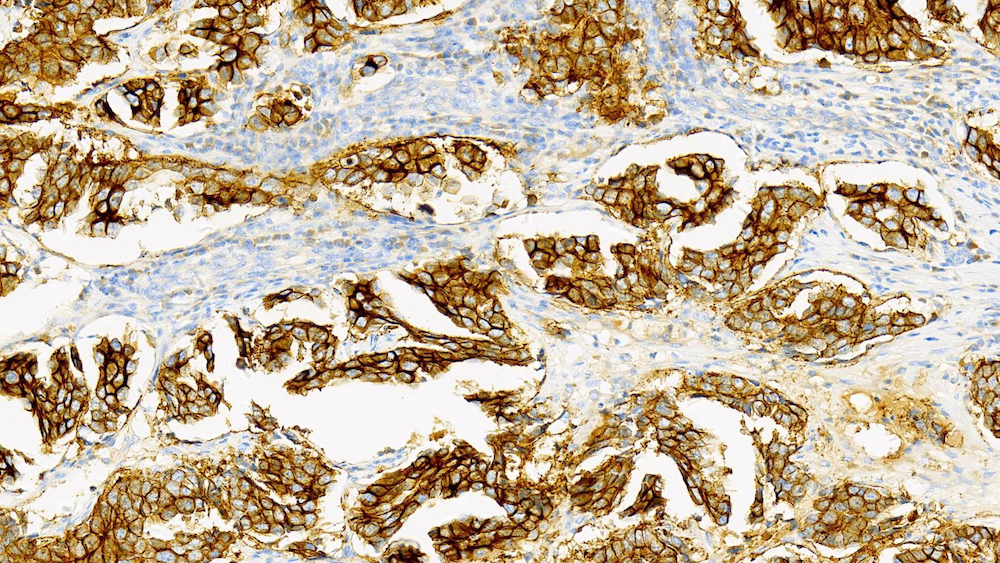}
\endminipage
\caption{FOV of breast cancer HER2 slides with negative (1+), equivocal (2+), and positive (3+) expressions}
\label{fig:her2example}
\end{figure}

The human epidermal growth factor receptor 2 gene HER2 is amplified in about 15\% to 20\% of invasive breast cancers \cite{her2amplify}. The HER2 amplification, associated with HER2 protein over-expression and a high rate of recurrence and mortality, is a vital predictor for patients with breast cancer \cite{her2amplify}. Therefore, the HER2 immunohistochemistry (IHC) test is used for a screening test as a routine practice for breast cancer in pathology, and a significant biomarker for providing effective therapies targeting the HER2 protein such as trastuzumab, lapatinib, and pertuzumab \cite{her22020}. Table \ref{table: her2guideline} shows the updated scoring criteria for breast HER2 IHC testing revised by the American Society of Clinical Oncology/College of American Pathologists (ASCO/CAP) in 2018 \cite{asco}. Moreover, cases of score 0 or 1+ are defined as HER2 negative and cases of score 3+ are defined as HER2 positive, while cases of 2+ are defined as equivocal which are required to be further assessed by situ hybridization (ISH) \cite{asco}. Overall, the scoring is based on the staining intensity and completeness of tumor cells. The examples of IHC negative, equivocal, and positive cases of the HER2 test are shown in figure \ref{fig:her2example} respectively.

Reading and scoring on tissue slides or whole slide images (WSIs) require pathologists to observe and calculate thousands even millions of tumor cells for classifying as positive, equivocal, or negative, so manually analysis takes considerable time and effort to evaluate, and is error-prone. Besides, the interobserver and intraobserver variations are inclined to find in the conventional assessment of HER2 slides. With the rapid development of artificial intelligence (AI) in diverse fields, there are several works unveiling solutions for these to use AI to assist pathologists with HER2 scoring. In general, the automated HER2 scoring is significant in clinical application by reducing the subjectivity and pathologists' workload. 

\begin{table}[h!]
\centering
\caption{\label{table: her2guideline}HER2 Scoring Guidelines\\ $ $}

\begin{tabular}{p{1.2cm}|p{1.9cm}|p{8.5cm}|p{3.15cm}}

\hline
\textbf{Score} & \textbf{Expression} & \textbf{Staining Pattern} & \textbf{Proportion} \\
\hline 
IHC 0 & Negative & No staining, or & $<10 \%$ of tumor cells \\
 & & faint/barely incomplete membrane staining & \\[.5ex]

IHC 1+ & Negative & Faint/barely perceptible incomplete membrane staining & $\geq 10 \%$ of tumor cells \\

IHC 2+ & Equivocal & Complete weak/moderate membrane
staining & $\geq 10 \%$ of tumor cells \\
 & & Complete and circumferential
membrane staining that is intense & $\leq 10 \%$ of tumor cells \\

IHC 3+ & Positive & Homogeneous, dark circumferential
membrane staining & $\geq 10 \%$ of tumor cells \\[.5ex]
\hline
\end{tabular}
\end{table}

Recently, the solutions for automated HER2 scoring are overall based on either conventional image processing or machine learning. For conventional techniques, for example, the HER2-CONNECT\textsuperscript \textregistered software uses a HER2 IHC algorithm to quantify the skeleton pattern of HER2 by measuring the connectivity and size distribution of stained membranes \cite{her2connect}. In the past, some of the works mainly employ approximations of area proportion of staining area as a measurement of scoring \cite{tradition1,tradition2,tradition3}. Recently, the characteristic curve defined for the stained area is used for some of the work to distinguish the different levels of staining, and therefore predict the HER2 scores (some machine learning classification algorithms may use for extracting features from characteristics curves) \cite{curve1,curve2}. 
However, these methods use disparate scoring rules in comparison with the HER2 scoring guidelines, such that it is unreasonable for clinical application.

On the other hand, machine learning (especially deep learning) methods are widely used for HER automated scoring. For example, the Her2Net is developed in  \cite{her2net} for identifying, segmenting, and classifying cell membranes and nuclei from WSI with few user interventions. Mostly, different deep convolutional neural network (CNN) architectures, and different pre- or post-processing techniques are used for segmentation and classification of HER2 scoring in WSI \cite{warwick, deep1, deep2, deep3}. Although these methods more or less achieve high accuracy of classification, the inexplicable result is the main open challenge for these deep learning methods, so that there is a long way for the deployment of such methods in real. 

In this paper, we proposed an automated scoring system of HER2 for quantitative analysis of the field of views (FOVs) under the microscope. Unlike the most of current work using approximations under the WSI, our scoring system strictly follows the recommended guidelines and provide real-time analysis and result visualizations for each FOV under the microscope while pathologist selecting several typical FOVs for diagnosing. The color intensity and completeness of staining regions around tumor cells are derived by various image processing techniques, and such processes make our system reasonable and interpretable for HER2 scoring. The proposed system also simulates the workflow of reading slides under the microscope and embeds interactive modification tools for selecting regions of interest (ROI) and adjusting parameters for intermediate mask generations based on tumor regions in each FOV, to improve the robustness, and furthest reduce pathologists workload in clinical application. 

Due to the variation of the staining kit, positive controls become necessary with every assay to ensure assay performance \cite{assay}. However, to our knowledge, there is almost no latest research considering positive controls for automated HER2 scoring, which may lead to varied results with different batches of staining. Therefore, instead of predicting HER2 score only based on WSI or FOVs, this paper will also introduce the color and intensity information of the positive control into our automated scoring process. In the future, this system will be used by pathologists in one of the first-class hospitals in China.

\section{HER2 Automated Scoring System}

\begin{figure}[H]
$$
\includegraphics[width=\linewidth]{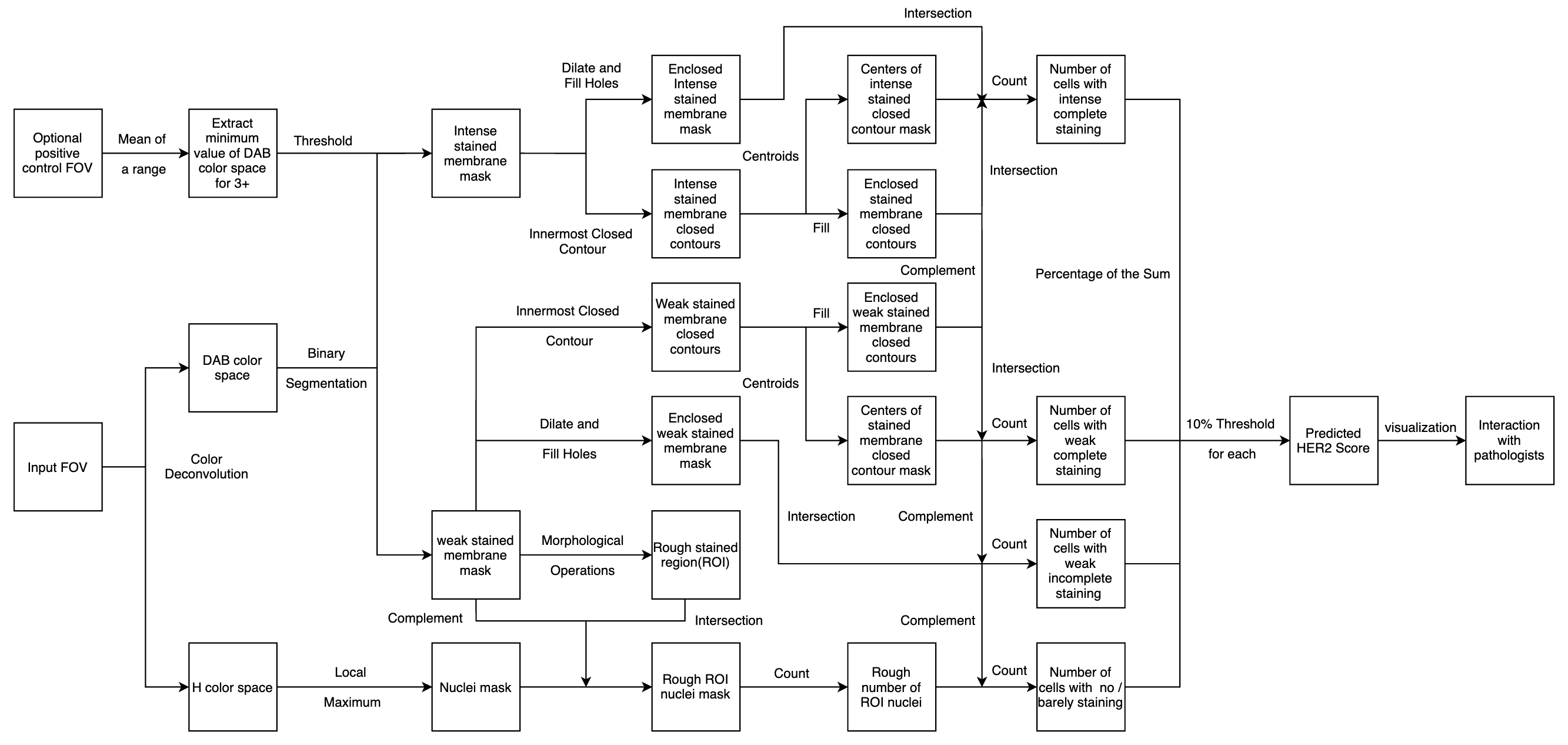} 
$$
\caption{HER2 Automated Scoring System workflow}
\label{workflow}
\end{figure}

Our method directly follows the HER2 scoring guidelines, and simulates the diagnostic workflow of the pathologists, considering the existence of the control positive FOV.  The algorithm workflow of the proposed HER2 automated scoring system is shown in figure \ref{workflow}.

\subsection{Materials}
The data used in this work is from WSIs scanned by a Motic scanner under the 40x magnification in People's Liberation Army General Hospital. The FOV in our system is $1920\times 1080$ under microscope with 40x and 20x objective lenses.

\subsection{Staining Membrane Segmentation}
The original input RGB image will be first separated from the brown IHC staining from the hematoxylin counterstaining using the color deconvolution method  \cite{HED}, noted as diaminobenzidine (DAB) channel in Haematoxylin-Eosin-DAB (HED) color space. Then, inspired by  \cite{microher2}, our method for staining membrane segmentation generates several masks representing different color and morphological information. Based on the HER2 scoring recommendations, the masks for weak or intense, and complete or incomplete staining expression are segmented for classifying tumor cells for each HER2 score class.

\begin{figure}[t] 
\begin{subfigure}{0.245\textwidth}
\includegraphics[width=\linewidth]{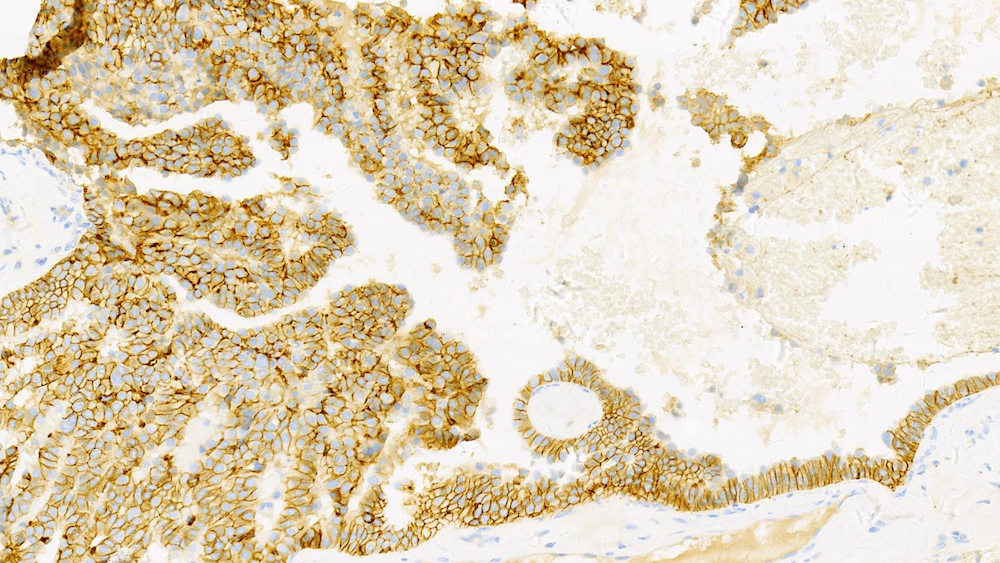}
\caption{Target FOV} \label{fig:a}
\end{subfigure}\hspace*{\fill}
\begin{subfigure}{0.245\textwidth}
\includegraphics[width=\linewidth]{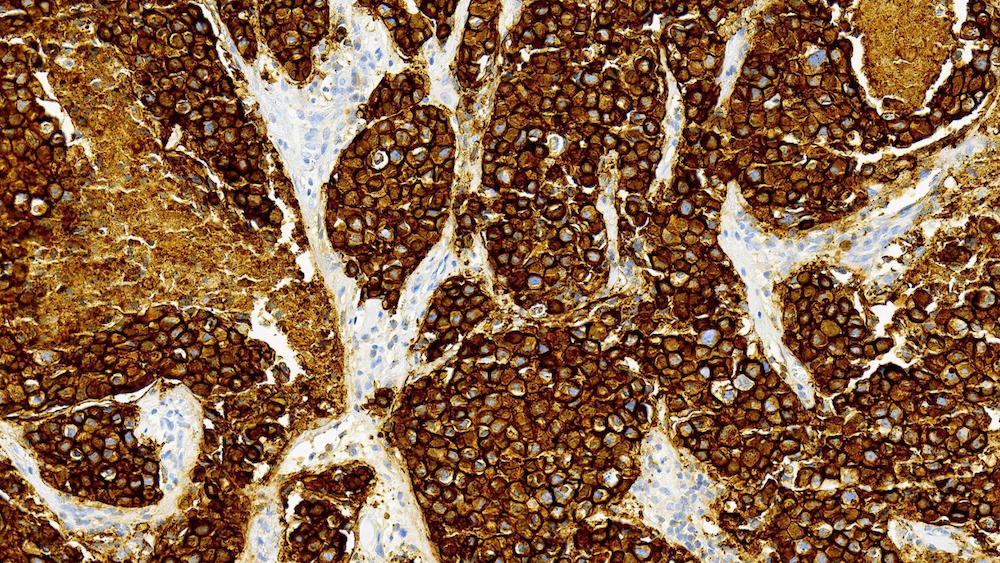}
\caption{Positive Control FOV} \label{fig:b}
\end{subfigure}\hspace*{\fill}
\begin{subfigure}{0.245\textwidth}
\includegraphics[width=\linewidth]{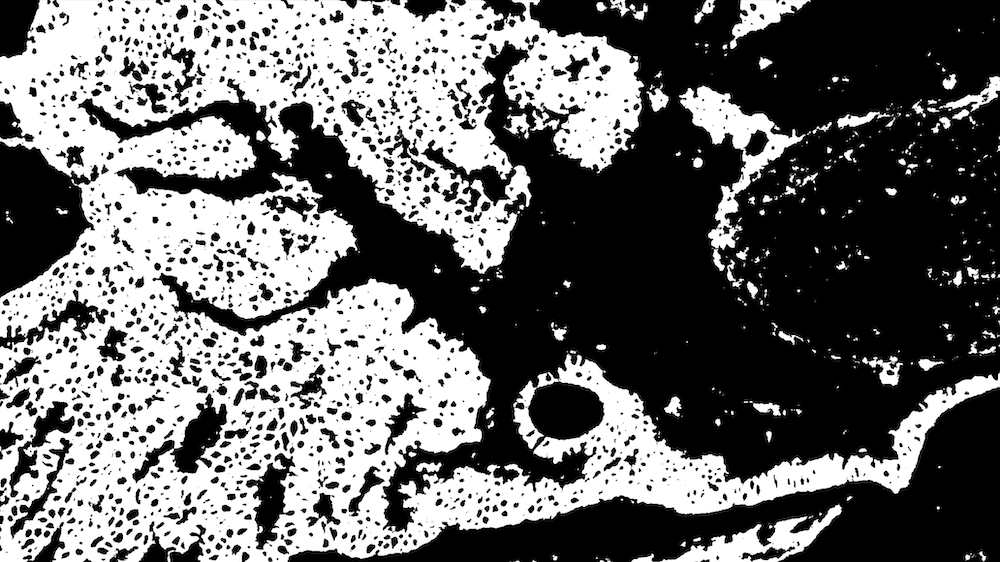}
\caption{$\mathbb{M}_{weak}$} \label{fig:c}
\end{subfigure}\hspace*{\fill}
\begin{subfigure}{0.245\textwidth}
\includegraphics[width=\linewidth]{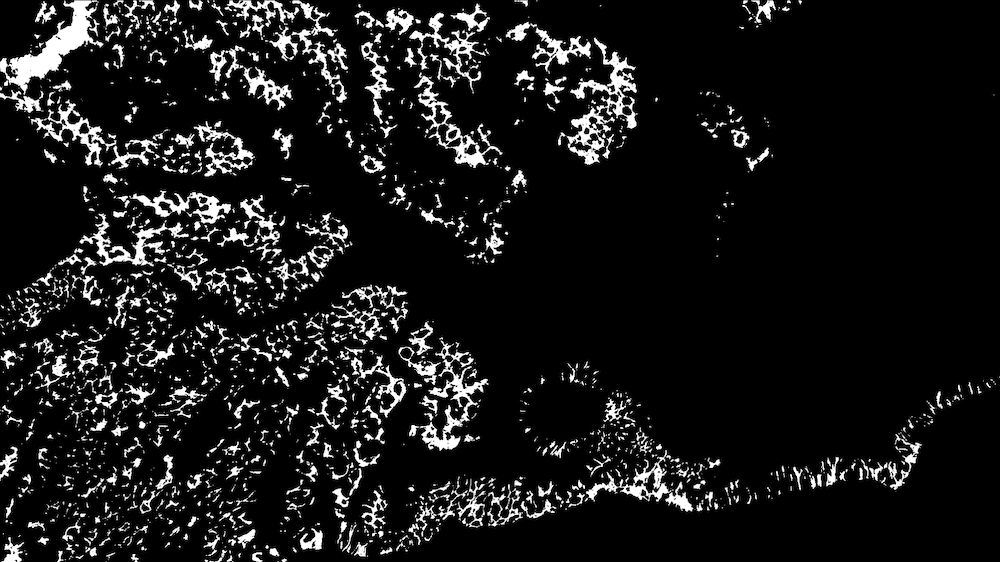}
\caption{$\mathbb{M}_{intense}$} \label{fig:d}
\end{subfigure}
\medskip
\begin{subfigure}{0.245\textwidth}
\includegraphics[width=\linewidth]{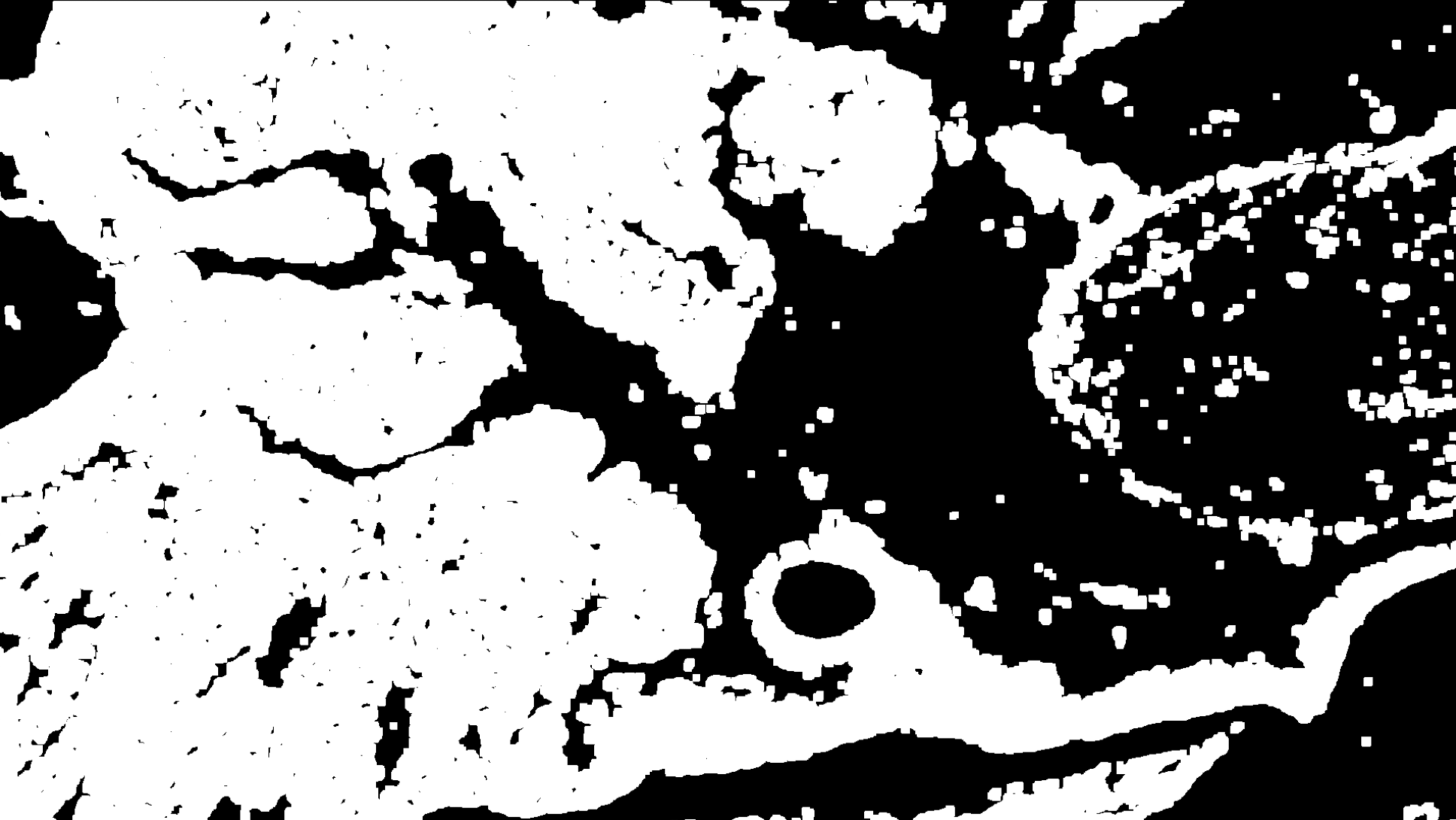}
\caption{$\mathbb{D}_{weak}$} \label{fig:e}
\end{subfigure}\hspace*{\fill}
\begin{subfigure}{0.245\textwidth}
\includegraphics[width=\linewidth]{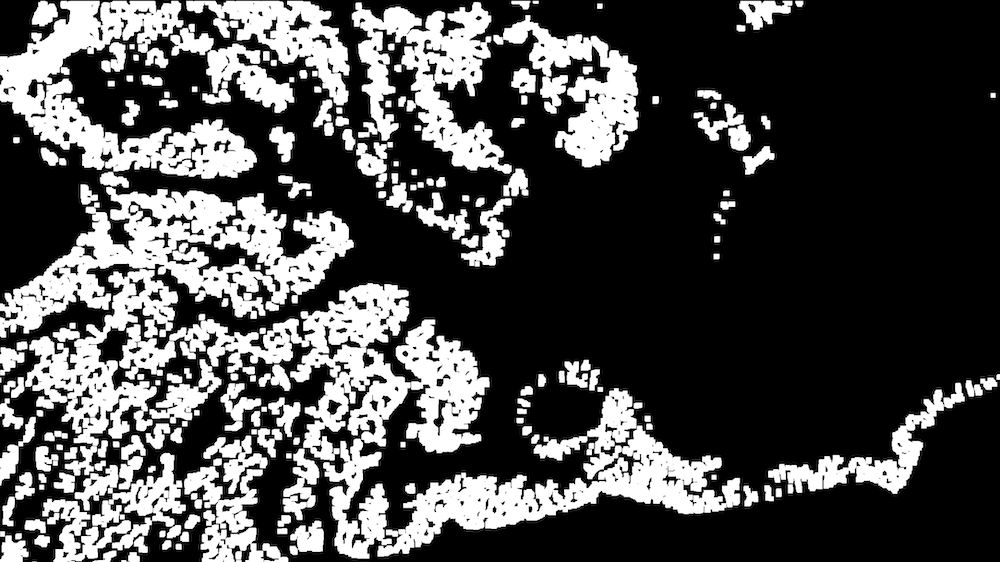}
\caption{$\mathbb{D}_{intense}$} \label{fig:f}
\end{subfigure}\hspace*{\fill}
\begin{subfigure}{0.245\textwidth}
\includegraphics[width=\linewidth]{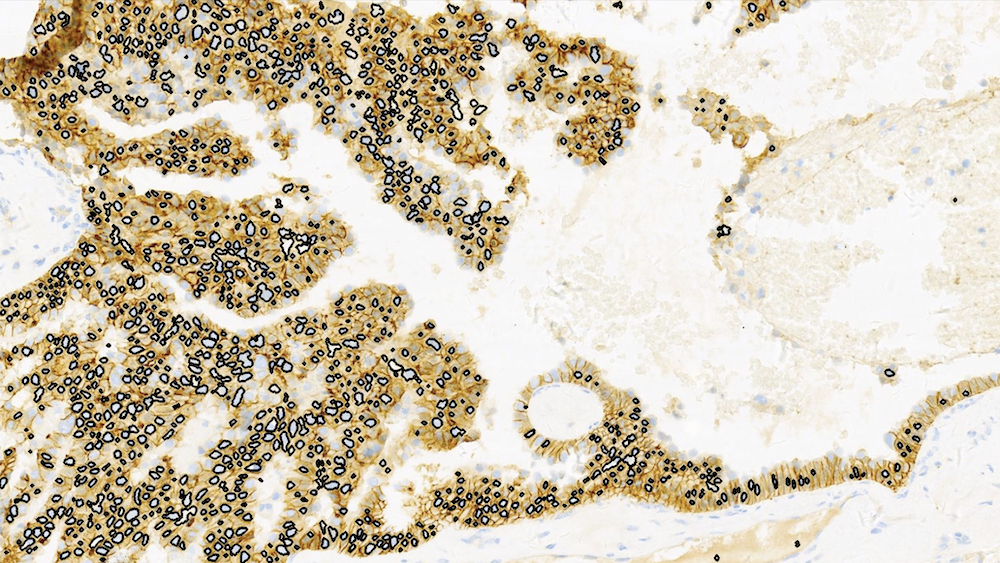}
\caption{$\mathbb{C}_{weak}$} \label{fig:g}
\end{subfigure}\hspace*{\fill}
\begin{subfigure}{0.245\textwidth}
\includegraphics[width=\linewidth]{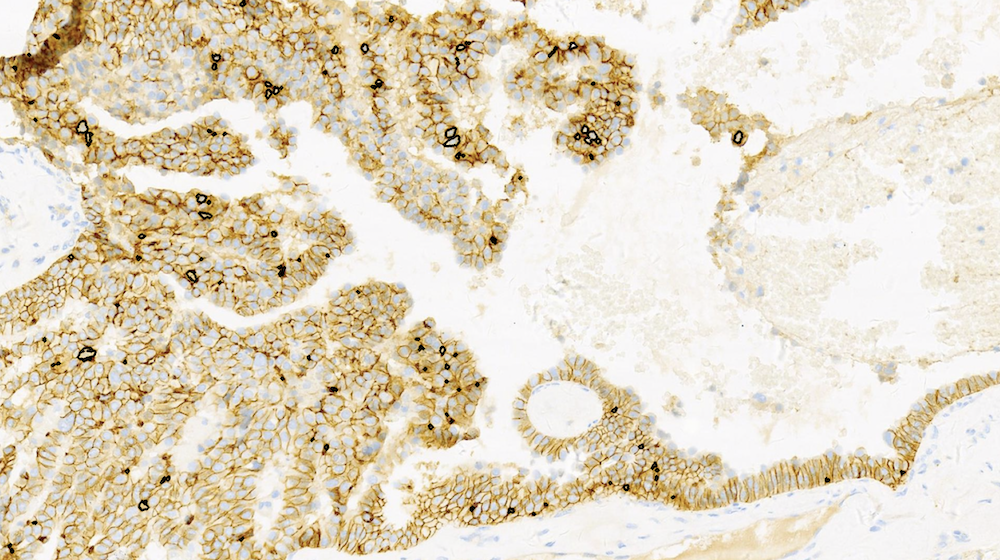}
\caption{$\mathbb{C}_{intense}$} \label{fig:h}
\end{subfigure}
\medskip
\begin{subfigure}{0.245\textwidth}
\includegraphics[width=\linewidth]{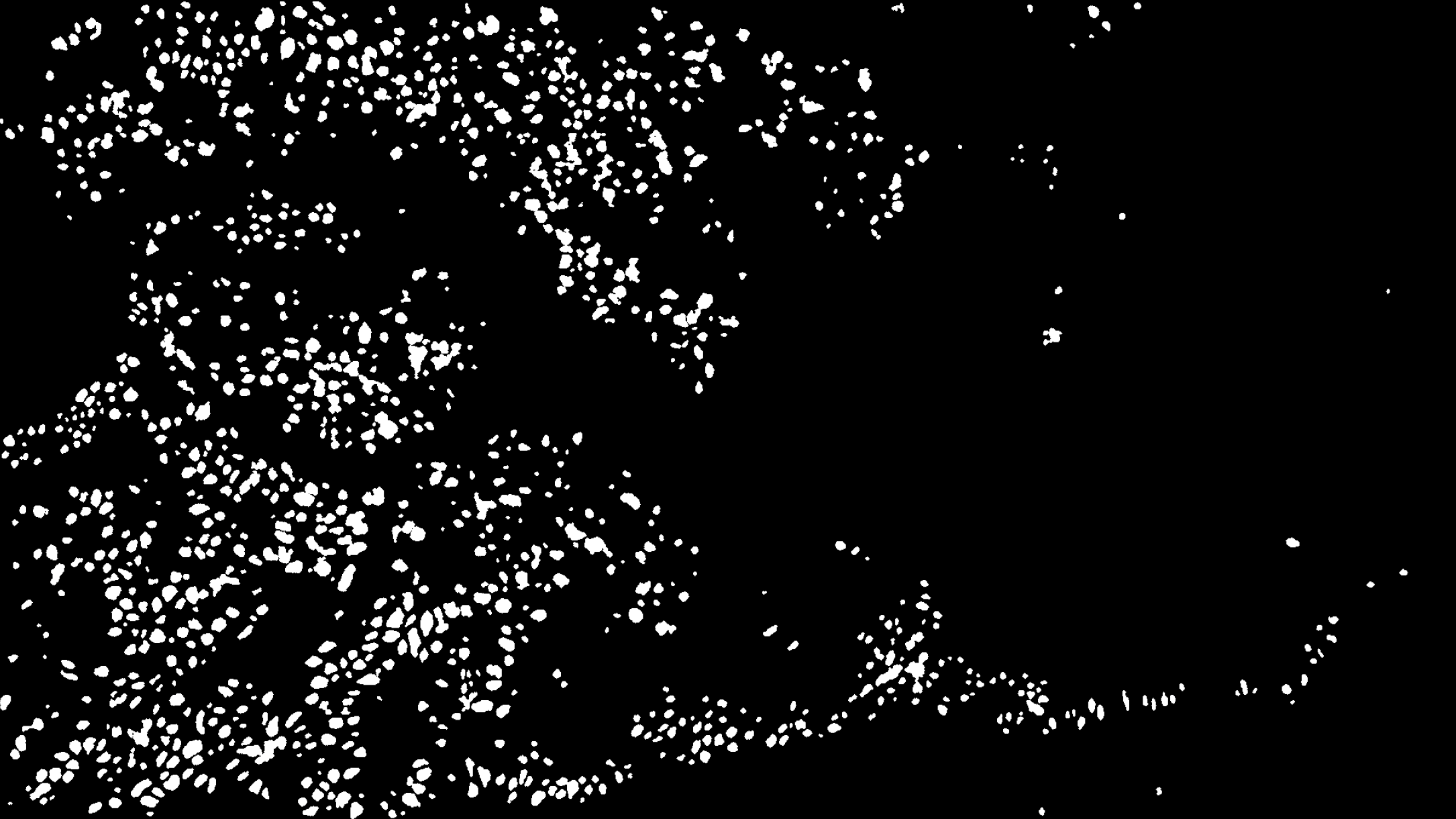}
\caption{$\mathbb{M}_{weak, complete}$} \label{fig:i}
\end{subfigure}\hspace*{\fill}
\begin{subfigure}{0.245\textwidth}
\includegraphics[width=\linewidth]{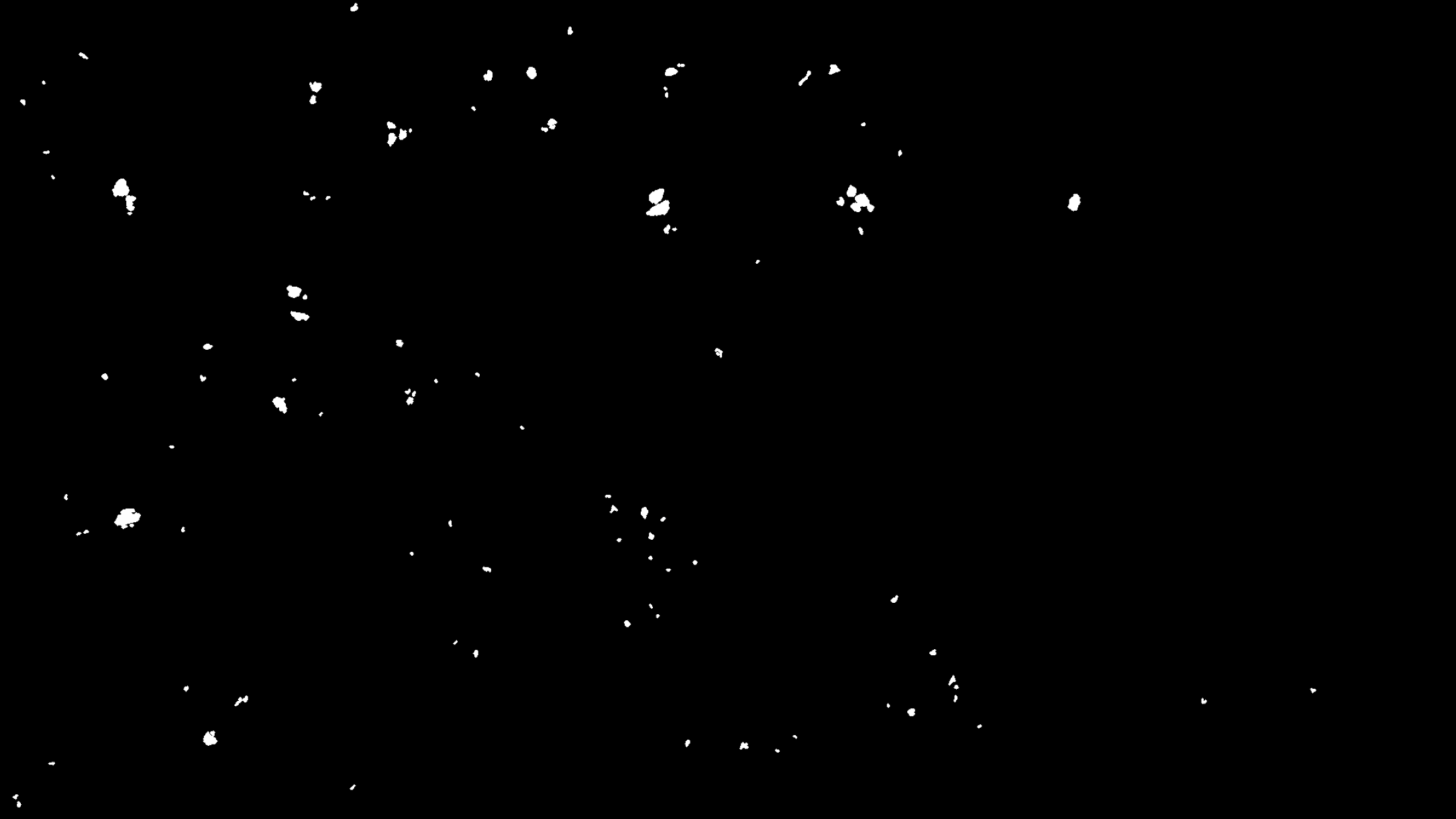}
\caption{$\mathbb{M}_{intense, complete}$} \label{fig:j}
\end{subfigure}\hspace*{\fill}
\begin{subfigure}{0.245\textwidth}
\includegraphics[width=\linewidth]{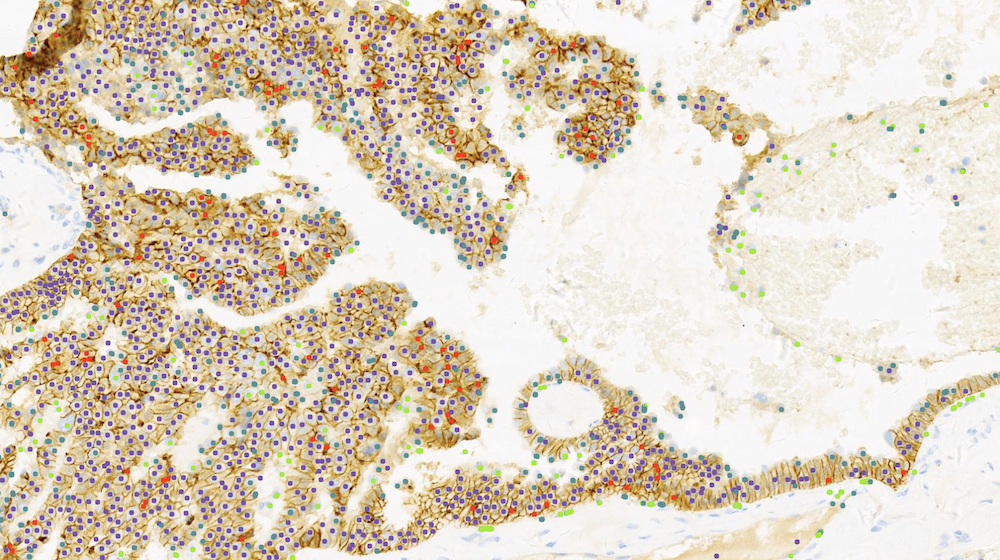}
\caption{Visualization} \label{fig:k}
\end{subfigure}\hspace*{\fill}
\begin{subfigure}{0.245\textwidth}
\includegraphics[width=\linewidth]{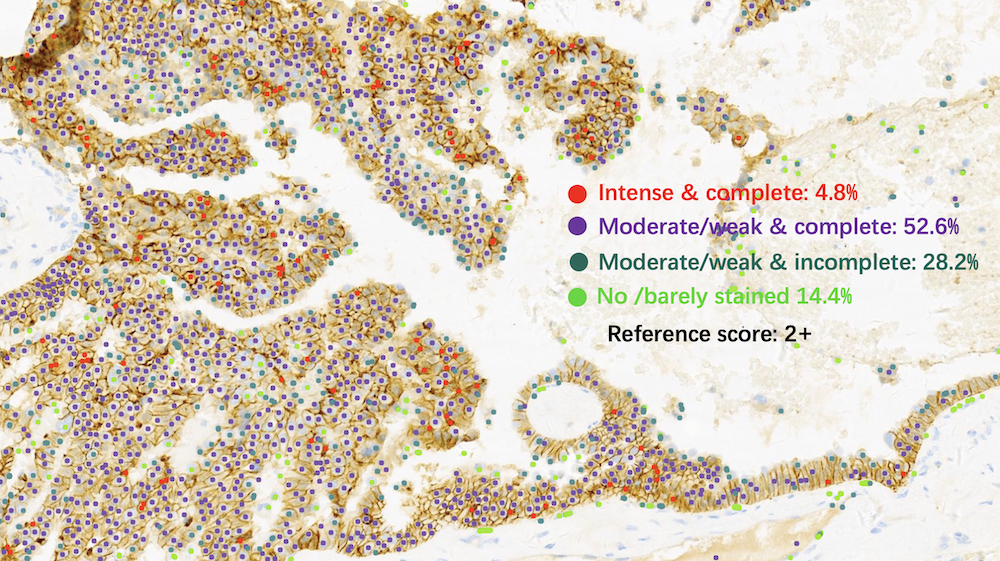}
\caption{Visualization with Analysis} \label{fig:l}
\end{subfigure}
\caption{The process of the proposed method. (a): Input target FOV; (b): positive control FOV; (c), (d), (e), (f), (g), (h), (i), (j): intermediate segmented masks; (k), (l): visualization and result} \label{fig:processes}
\end{figure}
\subsubsection{Threshold Extraction by Positive Control FOV}
We simply use a very small threshold to segment the DAB color space image to obtain the weakly stained mask, denoted as $\mathbb{M}_{weak}$ shown in figure \ref{fig:c}. The positive control FOV provides the color information of IHC positive (3+) where the intense staining is uniform and extensive, so the threshold we used to determine whether the target FOV staining is intense or not is directly corresponding with the DAB distribution of the positive control FOV. An example of positive control FOV is shown in figure \ref{fig:b}. Specifically, we use a mean of a range of DAB values by default as the threshold to obtain the intensely stained mask $\mathbb{M}_{intense}$, shown in figure \ref{fig:d}, for experiments to achieve the best prediction, and the rest of staining membrane segmentation processes are more or less based on these two masks. In consideration of the corner case during the real pathological diagnosis, we also add a panel for pathologists to select more specific regions to extract a more accurate intense staining threshold for convenience.

\subsubsection{Masks Extraction}\label{masks}
After the confirmation of the target FOV staining area by $\mathbb{M}_{weak}$ and $\mathbb{M}_{intense}$, using morphological dilation with a kernel of the tumor cell radius size provides masks $\mathbb{D}_{weak}$ (figure \ref{fig:e}), and $\mathbb{D}_{intense}$ (figure \ref{fig:f}) of cells with weak and intense membrane staining. These masks indicate the staining intensity information of each cell. 

To see the completeness of the membrane staining, we employ the border following algorithms in  \cite{contour} to retrieve contours from binary masks $\mathbb{M}_{weak}$ and $\mathbb{M}_{intense}$. The algorithm provides not only the border information but also the image topological information: for each $i^{th}$ contour $\mathbb{C}_i$, indexes of its first child contour and its parent counter of hierarchical levels can be retrieved as well. In our scenario, the complete membrane staining around each tumor cell is supposed to have the corresponding contour with no child but at least one parent contour to prevent including the approximation of incomplete membrane staining. Next, we filter out contours whose area is so small or large in comparison with the tumor cell, and fill contours to generate masks $\mathbb{M}_{weak, complete}$ and $\mathbb{M}_{intense, complete}$ for cells with complete weak and intense membrane staining respectively. The contours of weak and intense staining $\mathbb{C}_{weak}$ and $\mathbb{C}_{intense}$ visualized in the input FOV are shown in figure \ref{fig:g}, \ref{fig:h}, and the filled contour masks are shown in figure \ref{fig:i}, \ref{fig:j}. Note that $\mathbb{M}_{intense} \subseteq \mathbb{M}_{weak}$ by construction but preventing finding contours for the intense mask again by using $\mathbb{C}_{weak} \cap\mathbb{M}_{intense}$ is not feasible especially when the DAB values distribution of FOV pixels is closed to thresholds, due to the potential lack of the overview of membrane staining connectivity in each tumor region.

\subsection{Tumor Cell Segmentation and Counting}

In clinical application, the shape of tumor cells is more regular and larger than the other cell types (e.g. immune cells, stromal cells) and most of the tumor cells are stained in the tumor region in IHC HER2 slides. Then, to prevent expensive, challenging, severely scarce, and not perfectly guaranteed accurate annotations of different types of cells for classification by neural network models, our proposed cell detection method involves various image processing techniques. Notably, unlike the previous work, our method also takes the tumor cell that is not clearly shaped after being stained into account. As no pathologist could make sure that the stain is perfectly clear under the microscope every time, this process becomes necessary to increase the accuracy for cell detection and counting, particularly for dark staining in 3+ FOV (like the positive control FOV example shown in figure \ref{fig:b}). 

\subsubsection{Rough Cell Segmentation}

The Haematoxylin (H) channel in HED color space is firstly extracted for the segmentation of light blue cells. With a very small threshold, the cell mask is generated for segmentation. After the morphological opening of the cell mask, the clear nuclei can be obtained by finding the local maxima pixels of the foreground, where a small foreground proportion is properly set. As the purpose of segmentation is counting, no further segmentation algorithm needs to be employed such that over-segmentation can be adequately diminished. The binary mask for all clear nuclei is denoted as $\mathbb{M}_{nuclei, clear}$. Besides, dilation and erosion of stained areas (i.e. $\mathbb{M}_{weak}$) are used to determine the region of interest (ROI) for each selected FOV with tumor regions. To prevent the iteration of all pixels to observe the staining situation later, the coordinates of maxima point pixels are stored, denoted as the set $\mathbb{P}_{clear}$ with points $\mathbf{P}_{x, y}$ (i.e. $\mathbb{P}_{clear} = \{\mathbf{P}_{x, y} \mid \mathbb{M}_{nuclei, clear}(x, y) = 1, \mathbb{M}_{nuclei, clear}(x, y)\in \text{ROI}\}$). We will use $\mathbb{P}_{clear}$ for cells without staining or incomplete staining.

\subsubsection{Stained Cell Segmentation}

Almost all over- and under-segmentation cases of tumor cells appear in the stained area since the indeterminate connectivity of the membrane staining makes some cells not clear to observe and count. To solve this, we directly use the counters for intense or weak staining generated in section \ref{masks} to attain the location and the number of tumor cells with complete membrane staining. Specifically, centers of each counter are collected while generating $\mathbb{C}_{weak}$ and $\mathbb{C}_{intense}$. Then, points sets for coordinates of cells with weak complete staining and intense complete staining $\mathbb{P}_{weak, complete}$ and $\mathbb{P}_{intense, complete}$ are formed. Noted that $\mathbb{P}_{weak, complete} \subseteq \mathbb{P}_{intense, complete}$ now.

\subsubsection{Counting}

The number of cells assigned to each HER2 score category then can be obtained directly after segmentation. We have the numbers of cells with each HER2 score categories, denoted as $\mathcal{N}_{3+}$, $\mathcal{N}_{2+}$, $\mathcal{N}_{1+}$, and $\mathcal{N}_{0}$ ($|\cdot|$ for cardinality): 
\begin{alignat*}{2}
    &\mathcal{N}_{3+} &&= |\mathbb{P}_{intense, complete}|, \\
    &\mathcal{N}_{2+} &&= |\mathbb{P}_{weak, complete}\setminus \mathbb{M}_{intense, complete}|,  \\
    &\mathcal{N}_{1+} &&= |\mathbb{P}_{clear}\cap \mathbb{D}_{weak}\setminus(\mathbb{D}_{intense}\cup\mathbb{M}_{intense, complete} \cup \mathbb{M}_{weak, complete})|,\\
    & \mathcal{N}_{0} &&= |\mathbb{P}_{clear}\setminus(\mathbb{D}_{intense}\cup\mathbb{M}_{intense, complete} \cup \mathbb{M}_{weak, complete}\cup \mathbb{D}_{weak})|
\end{alignat*}

Note that as $\mathbb{C}_{intense}\neq \mathbb{C}_{weak}\cap \mathbb{M}_{intense}$ by our construction illustrated in section \ref{masks}, we can not use $\mathcal{N}_{2+} = |\mathbb{P}_{weak, complete}\setminus \mathbb{P}_{intense, complete}|$ but $|\mathbb{P}_{weak, complete}\setminus \mathbb{M}_{intense, complete}|$. This procedure then strictly follows the HER2 scoring guidelines and provides the quantitative counting for further classification.

\subsection{Scoring}
The HER2 scoring is based on the WSI in the application, so pathologists will select multiple representative target FOVs under the microscope and diagnose HER2 scores then. In our system, after selecting 5-10 FOVs including invasive breast cancer, quantitative analysis of each FOV and summarise of all FOVs then will be generated. Specifically, let $\mathcal{N}_{i} = \mathcal{N}_{3+_i} + \mathcal{N}_{2+_i} + \mathcal{N}_{1+_i}+ \mathcal{N}_{0_i}$ be the total number of tumor cells detected in $i^{th}$ target FOV by our method. In this way, for a set of FOVs, the predicted HER2 score is produced by the following rule:
\begin{enumerate}
    \item if $\displaystyle \frac{\sum \mathcal{N}_{3+_i}}{\sum \mathcal{N}_{i}} > 10\% $, the predicted HER2 score is 3+;
    \item else if $\displaystyle \frac{\sum \mathcal{N}_{2+_i}}{\sum \mathcal{N}_{i}} > 10\% $ or $\displaystyle \frac{\sum \mathcal{N}_{3+_i}}{\sum \mathcal{N}_{i}} < 10\% $, the predicted HER2 score is 2+;
    \item else if $\displaystyle \frac{\sum \mathcal{N}_{1+_i}}{\sum \mathcal{N}_{i}} > 10\% $, the predicted HER2 score is 1+;
    \item else, the predicted HER2 score is 0.
\end{enumerate}

\begin{figure}[H]
$$
\includegraphics[width=6.4in]{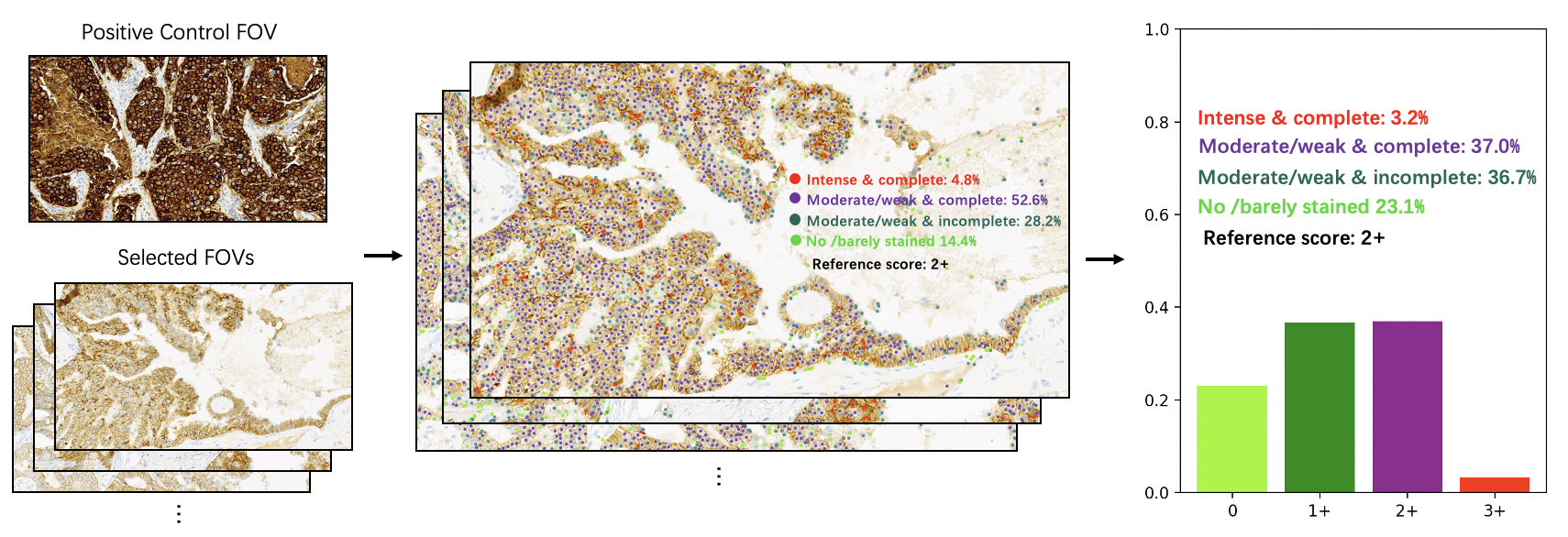} 
$$
\caption{Scoring Process and Visualization of Proposed System}
\label{scoring}
\end{figure}

Our method is also able to visualize and store each FOV's quantitative result in real-time while switching FOVs for the overall HER2 result. The cells with different staining intensities and completeness are varied by colors in the visualization. The process of scoring is shown in figure \ref{scoring}.

\section{Pathologists Interaction}

The pathological reading and diagnosis are exceedingly strict and tissues in FOVs under the microscope are intricate, so a onefold workflow seems impossible to satisfy all needs of the HER2 score prediction. Hence, the proposed system embedded in our \textit{Thorough Eye}\textsuperscript \textregistered platform provides several interactive and optional modifications for intermediate mask segmentation, as well as real-time visualizations of staining intensity and completeness in FOVs under the microscope. After these optional interactions, our method substantially reduces the workload of pathologists and keeps the explainable accuracy, achieving the main goals for the HER2 automated scoring method.

\subsection{Interactive Modifications}

While reading slides under a microscope by pathologists in real-time, it is difficult to guarantee the counting in each selected FOV is reasonable for invasive tumor regions. Therefore, we provide the selection of the ROI for processing the proposed counting and classification method, to improve the explainable accuracy. Besides, our platform has the panel to select representative staining in positive control FOV for threshold modification before extracting the intense staining mask $\mathbb{M}_{intense}$. For lightly stained FOVs, pathologists also have an option to select typical staining for modifying the threshold to segment $\mathbb{M}_{weak}$.

\subsection{Staining Visualizations}

We use different colors indicating cells with each category based on HER2 scoring guidelines, in order to keep the user interface (UI) clear for pathologists checking the membrane staining. Nevertheless, to assist pathologists to qualitatively visualize the membrane staining color intensity and completeness, we have a panel to show or hide these pieces of information for each FOV while generating masks by our system. 

\begin{figure}[H]
\begin{subfigure}{0.49\textwidth}
\centering
\includegraphics[width=0.95\linewidth]{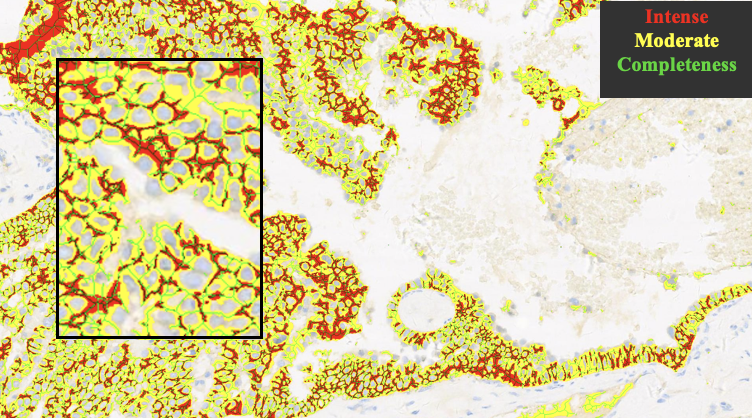}
\caption{intense and weak/moderate membrane staining visualization}
\label{fig:staining1}
\end{subfigure}
\begin{subfigure}{0.49\textwidth}
\centering
\includegraphics[width=0.95\linewidth]{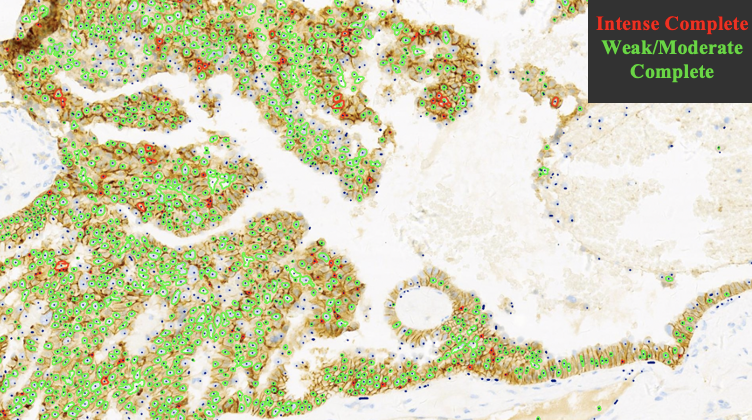}
\caption{complete membrane staining visualization}
\label{fig:staining2}
\end{subfigure}
\caption{Membrane Staining Intensity and Completeness Qualitative Visualization}
\label{fig:staining}
\end{figure}

Figure \ref{fig:staining1} indicates the optional visualization for intense membrane staining, which is colored red from $\mathbb{M}_{intense}$, and weak/moderate membrane staining, which is colored yellow from $\mathbb{M}_{weak} - \mathbb{M}_{intense}$. A thinning algorithm \cite{thin} is used here to perform the skeleton of the stained area for visualizing the completeness of membrane staining. Meanwhile, figure \ref{fig:staining2} indicates the optional visualization of complete and intense (red) or weak/moderate (green) membrane staining. The complete contours are the innermost contours of the membrane staining, derived from $\mathbb{C}_{intense}$ and $\mathbb{C}_{weak}$ for intense and weak/moderate staining respectively. 

\section{Conclusion and Future Work}

In this work, we present a quantitative analysis system for the automated scoring of IHC stained HER2 slides of breast cancer. Unlike the most related work that processes supervised learning in WSI, our method performs the explainable workflow for HER2 diagnosis strictly following the HER2 scoring guidelines under the microscope and produces quantitative and qualitative analysis in seconds empirically. It is worth noting that our system also takes the positive control into account, and allows pathologists to modify intermediate parameters in our algorithm, in order to improve the diagnostic robustness and accuracy, and substantially reduce the pathologists' workload while reading and diagnosis under the microscope. In the future, the proposed workflow of our system can be generated for more pathological assessments 
for other carcinomas. While deploying and using it in the first-class hospital, our system can have the chance to continuously improve and update after reading FOVs under the agreement with patients.

\section*{Acknowledgement}

This work is supported by \textit{Thorough Images}\textsuperscript \textregistered and pathologists in People's Liberation Army General Hospital.





\addcontentsline{toc}{section}{References}

\begin{thebibliography}{10}

\bibitem{her2amplify}
Dennis~J Slamon, Gary~M Clark, Steven~G Wong, Wendy~J Levin, Axel Ullrich, and
  William~L McGuire.
\newblock Human breast cancer: correlation of relapse and survival with
  amplification of the her-2/neu oncogene.
\newblock {\em science}, 235(4785):177--182, 1987.

\bibitem{her22020}
Soomin Ahn, Ji~Won Woo, Kyoungyul Lee, and So~Yeon Park.
\newblock Her2 status in breast cancer: changes in guidelines and complicating
  factors for interpretation.
\newblock {\em Journal of pathology and translational medicine}, 54(1):34,
  2020.

\bibitem{asco}
Antonio~C Wolff, M~Elizabeth~Hale Hammond, Kimberly~H Allison, Brittany~E
  Harvey, Pamela~B Mangu, John~MS Bartlett, Michael Bilous, Ian~O Ellis,
  Patrick Fitzgibbons, Wedad Hanna, et~al.
\newblock Human epidermal growth factor receptor 2 testing in breast cancer:
  American society of clinical oncology/college of american pathologists
  clinical practice guideline focused update.
\newblock {\em Archives of pathology \& laboratory medicine},
  142(11):1364--1382, 2018.

\bibitem{her2connect}
Anja Br{\"u}gmann, Mikkel Eld, Giedrius Lelkaitis, S{\o}ren Nielsen, Michael
  Grunkin, Johan~D Hansen, Niels~T Foged, and Mogens Vyberg.
\newblock Digital image analysis of membrane connectivity is a robust measure
  of her2 immunostains.
\newblock {\em Breast cancer research and treatment}, 132(1):41--49, 2012.

\bibitem{tradition1}
Yutaka Hatanaka, Kaoru Hashizume, Yuki Kamihara, Hitoshi Itoh, Hitoshi Tsuda,
  R~Yoshiyuki Osamura, and Yoichi Tani.
\newblock Quantitative immunohistochemical evaluation of her2/neu expression
  with herceptesttm in breast carcinoma by image analysis.
\newblock {\em Pathology international}, 51(1):33--36, 2001.

\bibitem{tradition2}
Hans-Anton Lehr, Timothy~W Jacobs, Hadi Yaziji, Stuart~J Schnitt, and Allen~M
  Gown.
\newblock Quantitative evaluation of her-2/neu status in breast cancer by
  fluorescence in situ hybridization and by immunohistochemistry with image
  analysis.
\newblock {\em American journal of clinical pathology}, 115(6):814--822, 2001.

\bibitem{tradition3}
Vilppu~J Tuominen, Teemu~T Tolonen, and Jorma Isola.
\newblock Immunomembrane: a publicly available web application for digital
  image analysis of her2 immunohistochemistry.
\newblock {\em Histopathology}, 60(5):758--767, 2012.

\bibitem{curve1}
Ramakrishnan Mukundan.
\newblock A robust algorithm for automated her2 scoring in breast cancer
  histology slides using characteristic curves.
\newblock In {\em Annual Conference on Medical Image Understanding and
  Analysis}, pages 386--397. Springer, 2017.

\bibitem{curve2}
Ramakrishnan Mukundan.
\newblock Analysis of image feature characteristics for automated scoring of
  her2 in histology slides.
\newblock {\em Journal of Imaging}, 5(3):35, 2019.

\bibitem{her2net}
Monjoy Saha and Chandan Chakraborty.
\newblock Her2net: A deep framework for semantic segmentation and
  classification of cell membranes and nuclei in breast cancer evaluation.
\newblock {\em IEEE Transactions on Image Processing}, 27(5):2189--2200, 2018.

\bibitem{warwick}
Talha Qaiser, Abhik Mukherjee, Chaitanya Reddy~Pb, Sai~D Munugoti, Vamsi
  Tallam, Tomi Pitk{\"a}aho, Taina Lehtim{\"a}ki, Thomas Naughton, Matt
  Berseth, An{\'\i}bal Pedraza, et~al.
\newblock Her 2 challenge contest: a detailed assessment of automated her 2
  scoring algorithms in whole slide images of breast cancer tissues.
\newblock {\em Histopathology}, 72(2):227--238, 2018.

\bibitem{deep1}
Tomi Pitk{\"a}aho, Taina~M Lehtim{\"a}ki, John McDonald, and Thomas~J Naughton.
\newblock Classifying her2 breast cancer cell samples using deep learning.
\newblock In {\em Proc. Irish Mach. Vis. Image Process. Conf.}, pages 1--104,
  2016.

\bibitem{deep2}
Michel~E Vandenberghe, Marietta~LJ Scott, Paul~W Scorer, Magnus S{\"o}derberg,
  Denis Balcerzak, and Craig Barker.
\newblock Relevance of deep learning to facilitate the diagnosis of her2 status
  in breast cancer.
\newblock {\em Scientific reports}, 7(1):1--11, 2017.

\bibitem{deep3}
Fariba~Damband Khameneh, Salar Razavi, and Mustafa Kamasak.
\newblock Automated segmentation of cell membranes to evaluate her2 status in
  whole slide images using a modified deep learning network.
\newblock {\em Computers in biology and medicine}, 110:164--174, 2019.

\bibitem{assay}
David~G Hicks and Linda Schiffhauer.
\newblock Standardized assessment of the her2 status in breast cancer by
  immunohistochemistry.
\newblock {\em Laboratory Medicine}, 42(8):459--467, 2011.

\bibitem{HED}
Arnout~C Ruifrok, Dennis~A Johnston, et~al.
\newblock Quantification of histochemical staining by color deconvolution.
\newblock {\em Analytical and quantitative cytology and histology},
  23(4):291--299, 2001.

\bibitem{microher2}
Jun Zhang, Kuan Tian, Pei Dong, Haocheng Shen, Kezhou Yan, Jianhua Yao, Junzhou
  Huang, and Xiao Han.
\newblock Microscope based her2 scoring system.
\newblock {\em arXiv preprint arXiv:2009.06816}, 2020.

\bibitem{contour}
Satoshi Suzuki et~al.
\newblock Topological structural analysis of digitized binary images by border
  following.
\newblock {\em Computer vision, graphics, and image processing}, 30(1):32--46,
  1985.

\bibitem{thin}
Zicheng Guo and Richard~W Hall.
\newblock Parallel thinning with two-subiteration algorithms.
\newblock {\em Communications of the ACM}, 32(3):359--373, 1989.

\end{thebibliography}

\end{document}